\documentclass[11pt]{article}
\begin{document}
\begin{titlepage}
\title{Comment on `Quantum-Anti-Zeno Paradox'}
\author{Lajos Di\'osi\footnote{E-mail: diosi@rmki.kfki.hu}\\
Research Institute for Particle and Nuclear Physics}
\maketitle
\begin{abstract}
One can reduce the involved derivation of Balachandran and Roy of 
their `anti-Zeno' effect [Phys.Rev.Lett. {\bf 84}, 4019 (2000)]
to the derivation of standard Zeno-effect. The mechanism of what the 
authors call `anti-Zeno' effect is a dynamic version of Zeno effect.
\end{abstract}

In a recent Letter [1], the well-known quantum-Zeno paradox is 
re-derived. 
The authors obtain the correct effect in a complicated form. 
Furthermore, they suggest as if the quantum-Zeno 
effect were only related to the continuous measurement of a 
{\it constant} projection operator $E$. The Letter then exposes 
the issue of continuous measurement of the time-dependent 
projection $E_t=U_tEU_t^\dagger$, derives equations of state 
evolution, and interpretes the result as anti-Zeno paradox. 
The Letter claims that the time-dependent case has a new, even 
an opposite, physical consequence in comparison with the case of 
constant $E$. This is a misleading suggestion, I am afraid.
The central purpose of my Comment is to correct the author's 
interpretation. I think the Zeno-mechanism itself is the same 
whether $E$ is time-dependent or not. This is a most productive 
starting point. I derive the Letter's main result from the equations 
of common quantum-Zeno effect (where $E=const.$), using elementary 
transformations. The elegance of this method as well as the 
transparency of the resulting equation will be convincing.
 
The standard quantum-Zeno effect means in fact the constancy of the 
measured value ($1$ or $0$) of the continuously measured projector 
$E$. The constancy of the quantum state follows from the constancy
of $E$. If $E$ depends on time in a smooth enough way, the standard 
Zeno effect survives perfectly: The value of a continuously 
measured time-dependent projector $E_t$ remains constant ($0$ or $1$), 
independently of the self-dynamics of the system. Correspondingly, 
Zeno effect means in its full generality that the quantum state 
$\psi_t$ remains confined in the subspaces belonging to $E_t$ for
all time $t$ during the continuous measurement. If $E_t$ projects 
on a one-dimensional subspace then, provided the first measurement 
gives $1$, i.e. $\psi_0\psi_0^\dagger=E_0$, the 
continuous measurement will be dragging the state $\psi_t$ with 
itself and, independently of the Hamiltonian, the identity 
$\psi_t\psi_t^\dagger=E_t$ remains valid during the continuous
measurement. This dragging-mechanism (known already to von Neumann 
and to others [2]  
as pointed out in Refs.~[3,4]) is the veriest quantum-Zeno 
mechanism. It is misleading to interprete it as a new anti-Zeno effect. 
[The author's twist to the old metafore is a bit artificial. The 
watched kettle ``boils'' {\it iff} the path  $\{E_t; t\in(0,T)\}$ 
reaches a boiling state. This is not necessarily sure for ``most 
ways of watching''.]

To support my viewpoint, I present an elementary derivation of 
the Letter's result, starting from the equations of standard Zeno
effect. So, consider a system whose unitary evolution 
$\dot\psi_t=-iH_t\psi_t$ is governed by an arbitrarily given 
Hamiltonian (maybe of smooth time-dependence). Let the 
{\it constant} projector $E$ be continuously measured and let us 
assume that the initial measured value is $1$, {\it i.e.}, 
$\psi_0=E\psi_0$. Then, the state will 
unitarily evolve with the effective Hamiltonian:
\begin{equation}\label{H}
\dot\psi_t=-iEH_tE\psi_t~.
\end{equation}
As a consequence, the identity
$E\psi_t=\psi_t$ 
will hold during the measurement; this is the standard Zeno effect.
Eq.~(\ref{H}), which I take here granted, follows from common
treatment of Zeno effect [5]. A smooth time-dependence of the 
Hamiltonian makes no difference [4].
Now we switch to the general case. Following the Letter, the smooth 
time-dependence of the measured projector $E_t$ will be described by 
unitary rotations $E_t=U_tEU_t^\dagger$, we set $U_0=1$. Note that  
the rotations $U_t$ are not unique while the projectors $E_t$ are. 
It is trivial to inspect that this case goes back to the standard one
in a frame rotated by $U_t$. Accordingly, I transform the state 
space as well as the observables: 
$\psi_t\rightarrow U_t^\dagger\psi_t$,
$O_t\rightarrow U_t^\dagger O_t U_t$.
Also the self-Hamiltonian  transforms: 
$H_t\rightarrow U_t^\dagger H_t U_t + i\dot U_t^\dagger U_t$. 
Obviously, the projector to be observed becomes the constant $E$ 
itself. We can therefore apply the Eq.~(\ref{H}) of 
standard quantum-Zeno. Then we return from the rotating frame to the
original one, this equation takes the following ultimate form:
\begin{equation}\label{Ht}
\dot\psi_t=-iE_tH_tE_t\psi_t+[\dot E_t,E_t]\psi_t~,
\end{equation}
leading to the general Zeno effect:
$E_t\psi_t=\psi_t$. 
We see that the evolution (\ref{Ht}) is always unitary since the 
effective Hamiltonian $E_tH_tE_t+i[\dot E_t,E_t]$ is hermitian. This
Hamiltonian evolution applies to the general case when the state
is described by density matrix. Another advantage of the above form 
is that, contrary to the Letter's Eq.(34), the unphysical unitary 
operators $U_t$ do not at all appear in it. Only the projectors $E_t$ 
do. The simplicity of Eq.~(\ref{Ht}) is remarkable as compared to the 
Letter's Eq.~(34). 

It is worthwile to add that the term `anti-Zeno' had already been
reserved for environmental {\it acceleration} of quantum transitions,
see,  e.g., Ref.[6] and references therein.

This work is supported by the Hungarian OTKA Grant T032640.

\vskip .5cm

[1] A.P. Balachandran and S.M. Roy, Phys. Rev. Lett. {\bf 84}, 4019 (2000).

[2] J. von Neumann, Matematische Grundlagen der Quantenmechanik 
	      (Springer, Berlin, 1932);
    Y. Aharonov and M. Vardi, Phys. Rev. {\bf D21}, 2235 (1980).

[3] D. Giulini, E. Joos, C. Kiefer, J.Kupsch, I.-O. Stamatescu and  
              H.D. Zeh: {\it Decoherence and the Appearance of a Classical
	      World in Quantum Theory\/} (Springer, Berlin, 1996).

[4] A. Peres, {\it Quantum Theory: Concepts and Methods\/} (Kluwer, 
	      Dordrecht, 1993). 	

[5] To verify the effective Hamiltonian $EH_tE$, consider the infinitesimal 
    unitary evolution of $\psi_t=E\psi_t$ broken by the measurement of $E$ 
    at time $t+dt$ which amounts in $Ee^{-idtH_t}\psi_t$. This we can 
    re-write as $e^{-idtEH_tE}\psi_t$.

[6] O.V. Prezhdo, Phys. Rev. Lett. {\bf 85}, 4413 (2000).
 
\end{titlepage}
\end{document}